\documentclass[sigconf]{acmart}
\usepackage[utf8]{inputenc}
\usepackage{multirow}
\usepackage{multicol}
\usepackage{tabularx} 
\usepackage[ruled,vlined,linesnumbered]{algorithm2e}
\newcolumntype{L}{>{\raggedright\arraybackslash}X}
\usepackage{amsmath}
\usepackage{amsthm}
\usepackage{siunitx}
\usepackage{makecell}
\usepackage{mathtools}
\usepackage{wrapfig}
\usepackage{subcaption}

\AtBeginDocument{%
  }

\setcopyright{none}
\renewcommand\footnotetextcopyrightpermission[1]{}

\begin{document}

\title{HOB: A Holistically Optimized Bidding Strategy under Heterogeneous Bidding Environments}

% \title {Holistic Optimization for Bidding in Mixed Bidding Mechanisms}
\settopmatter{authorsperrow=3} 
\author{Qi Li}
\authornote{These authors contributed equally to this research.}
\authornote{Corresponding author.} % Note for the 
\affiliation{%
  \institution{Alibaba Group}
  \city{Beijing}
  \country{China}
}
\email{luyuan.lq@alibaba-inc.com}

\author{Wendong Huang}
\authornotemark[1]
\affiliation{%
  \institution{Tsinghua University}
  \city{Beijing}
  \country{China}
}
\email{hwd23@mails.tsinghua.edu.cn}

\author{Qichen Ye}
\authornotemark[1]
\affiliation{%
  \institution{Alibaba Group}
  \city{Beijing}
  \country{China}
}
\email{yeqichen.yqc@alibaba-inc.com}

\author{Wutong Xu}
\author{Cheems Wang}
\affiliation{%
  \institution{Tsinghua University}
  \city{Beijing}
  \country{China}
}
% \email{xwt22@mails.tsinghua.edu.cn}
% \email{hhq123go@gmail.com}

\author{Wei Yuan}
\author{Miao Xu}
\author{Zhiyu Mou}
\author{Guan Wang}
\affiliation{%
  \institution{Alibaba Group}
  \city{Beijing}
  \country{China}
}
% \email{{jiarui.yw,xumiao.xm}@alibaba-inc.com}
% \email{xumiao.xm@alibaba-inc.com}
% \email{mouzhiyu.mzy@alibaba-inc.com}
% \email{fanghan.wg@alibaba-inc.com}

\author{Rongquan Bai}
\authornotemark[2]
\author{Chuan Yu}
\author{Jian Xu}
\affiliation{%
  \institution{Alibaba Group}
  \city{Beijing} % Consistent spelling
  \country{China}
}
\email{rongquan.br@alibaba-inc.com}
% \email{yuchuan.yc@alibaba-inc.com}
% \email{xiyu.xj@alibaba-inc.com}

\renewcommand{\shortauthors}{Qi Li, et al.}

\begin{abstract}
Optimizing a single advertising campaign across heterogeneous channels is a central challenge in industrial autobidding. Auction mechanisms vary across channels in ranking rules (pure eCPM vs. UE-augmented scoring), pricing formats (first- vs. second-price), and bidding conventions (uniform vs. non-uniform), while advertisers impose shared campaign-level constraints.
We propose HOB, which makes marginal cost (MC) computable and alignable across heterogeneous channels, especially for first-price auctions (FPA) with organic–paid coexistence, where existing bidding formulations do not yield a practical aligned MC form. At the global level, HOB derives channel-specific MC forms and coordinates disparate channels through a shared MC target. At the local level, HOB models free-win probability and winning-price uncertainty with a zero-inflated exponential distribution, yielding an efficient surplus-optimal bidding strategy for non-uniform first-price auctions. We show that any interior optimum satisfies MC equalization across channels. Experiments on a controlled offline benchmark, industrial log replay, and large-scale online A/B tests demonstrate that HOB consistently delivers significant performance gains. Deployed on a large-scale commercial DSP, HOB delivers a 3.0\% lift in GMV while maintaining return on advertising spend (ROAS) constraints.

% Optimizing a single advertising campaign across heterogeneous channels is a central challenge for industrial autobidding. Channels differ in auction formats, bidding rules, and traffic characteristics, making uniform bidding suboptimal under global advertiser constraints, such as budget and ROAS.
% Meanwhile, advertisers are shifting from buying only traffic ranked purely by eCPM to platform-wide solutions, where bids act as an eCPM-based boost on top of each request’s user-experience (UE) score. In FPA channels, high-UE requests could win with little or even zero payment, so overbidding wastes budget and reduces overall efficiency.
% To address these issues, we propose HOB, a unified bidding framework that decomposes the problem into two coupled components: (i) cross-channel coordination via marginal cost alignment, and (ii) request-level bidding for FPA that models free-win probability and winning-price uncertainty. We prove that marginal-cost equalization characterizes the optimum under standard regularity assumptions, and demonstrate significant GMV gains in large-scale offline evaluation and online A/B tests. The framework has been productionized at scale, serving millions of advertisers and resulting in a 3.0\% increase in GMV.

\end{abstract}




\maketitle
% \noindent \textbf{Presentation Preference:} We prefer a talk presentation.
% \vspace{1em}
% \pagestyle{plain} % 必须放在 \maketitle 之后
\section{Introduction}
\begin{figure}[t]
\includegraphics[width=\linewidth]{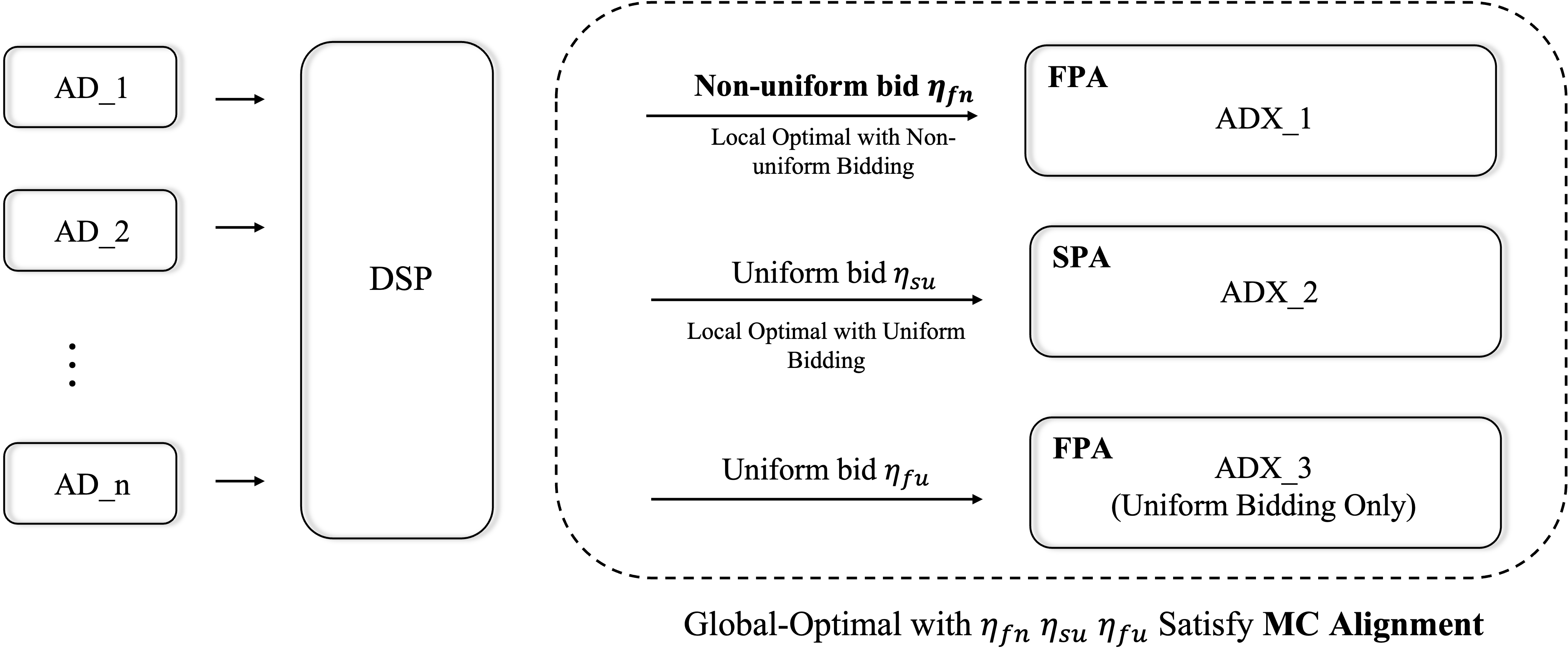}
\Description{A block diagram showing a DSP module receiving multiple AD inputs and distributing them into three heterogeneous bidding environments: ADX_1 using FPA with non-uniform bidding, ADX_2 using SPA with uniform bidding, and ADX_3 using FPA with uniform bidding. The system achieves global optimality where the bidding strategies satisfy MC alignment.}
\caption{Illustration of heterogeneous bidding environments.}
\label{Fig:intro_fig}
% \vspace{-20pt}
\end{figure}
% \begin{figure}[h]
%   \centering
%   \includegraphics[
%         % trim={0cm 2cm 0cm 2cm}, % 定义裁剪的边界
%         % clip,
%         width=\linewidth
%   ]{figures/hbo_intro.png}
%   \caption{Illustration of the procedure. Our work focuses on the design of holistically optimized bidding strategy.}
%   \Description{intro_fig}
% \end{figure}
% part 1: discuss SPA, FPA, optimal bidding
Automated advertising systems have become a major revenue source for modern internet platforms~\cite{mou2022sustainable,geng2021automated,borissov2010automated,jiang2025automated}. 
Before 2018, most platforms adopted second-price auctions (SPA) with uniform bidding, where each campaign bid as $bid_j=\eta \cdot Value_j$ by tuning a single scalar $\eta$, and the winner paid the second-highest bid~\cite{cramton2004competitive,zhao2023uniform}. Here, 
$Value_j$ denotes the expected advertiser value per impression, defined toward different objectives such as expected clicks, conversions or GMV.
The industry-wide shift to first-price auction (FPA) has fundamentally changed this paradigm~\cite{SPA_prob,bergemann2017first,paes2020competitive,deng2022Efficiency}. 
FPA has gained increasing adoption because it offers higher revenue potential for publishers and reduces concerns over discriminatory treatment induced by personalized reserve prices. 
Since bidders now pay their own bids, optimal bidding becomes request-level and inherently non-uniform~\cite{2021Towards}. 
Prior work has therefore studied non-uniform bidding methods that account for the competitive landscape~\cite{shading1, shading2, shading3}. 

Meanwhile, modern DSPs increasingly provide auto-bidding services that enable a single campaign to bid across multiple heterogeneous channels, as illustrated in Figure~\ref{Fig:intro_fig}.
However, existing studies predominantly focus on per-channel optimization, leaving the  cross-channel coordination across heterogeneous environments insufficiently explored.
This challenge is further complicated by the integration of paid signals into allocations that were previously organic~\cite{li2025beyond}. 
In many platform-wide marketing solutions, allocation is determined by a hybrid score combining user experience (UE) and an ad's effective cost per mille (eCPM), i.e.,
$FinalScore = UE + \alpha \cdot eCPM$. 
Under such a mechanism, an ad with sufficiently high \textit{UE} may still win exposure even with a zero bid.

We therefore target two coupled challenges: (i) cross-channel
coordination under shared budget/ROAS constraints, and (ii) within-
channel optimality in FPA with organic–paid coexistence.
% HOB solves heterogeneous-channel autobidding by making marginal cost computable and alignable across channels, especially for FPA with organic–paid coexistence where existing bidding formulations do not yield a practical aligned MC form.
Our contributions are:
(i) \textbf{Global optimality under heterogeneous channels.} We formulate multi-channel autobidding with shared budget and ROAS constraints, and provide a practical update rule to adjust channel bidding parameters toward this equilibrium;
(ii) \textbf{Optimal FPA bidding under organic–paid coexistence.} We derive request-level surplus-optimal bids by modeling free wins and conditional winning-price distributions;
(iii) \textbf{Production-scale effectiveness.} Offline experiments and large-scale online A/B tests show significant GMV gains under the same constraints, and we report a scalable production deployment in a commercial system.
\section{Related Work}
\label{related work}

\subsection{Auto Bidding}
\label{sec:auto_bidding}
The core task of an automated bidding system is to optimize an advertiser's objectives under various constraints~\cite{mou2022sustainable,geng2021automated,borissov2010automated,jiang2025automated,he2021unified}. Existing work broadly falls into two lines. The first studies \emph{how to bid}, formulating bidding as sequential decision-making and solving it with PID control, online linear programming, reinforcement learning or conditional action generation~\cite{cai2017Real,jin2018Real,mou2022sustainable, chen2021decision, guo2024generative}.
The second re-examines \emph{what to bid for}. Instead of optimizing immediate revenue per exposure, uplift-based methods optimize incremental value to better capture causal effects~\cite{hao2020dynamic,waisman2025Online,lewis2022incrementalitybiddingattribution,10.5555/3015812.3015909,moriwaki2020unbiasedliftbasedbidding}.
A common limitation is that most studies assume a simplified market, neglecting the complexities of modern advertising ecosystems that feature heterogeneous bidding environments.

\subsection{Bidding Across Multiple Channels}
\label{sec:multi_channel}
Cross-channel bid optimization has attracted growing attention. Prior work studies utility maximization under budget constraints~\cite{susan2023multi} and value maximization with additional target-ROAS constraints~\cite{deng2023multi}. Recent theory further suggests that, ignoring budget exhaustion, the optimal strategy equalizes marginal cost (MC) across channels~\cite{aggarwal2024platformcompetitionautobiddingworld}. ~\cite{aggarwal2025multiplatform} study the same MC equalization principle under abstract value/cost functions and propose efficient search algorithms with provable guarantees. However, our work operates at an abstraction level: we derive explicit MC forms for each auction type and provide a deployable online alignment mechanism.

\subsection{Bid Shading}
\label{sec:bid_shading}
Bid shading is a standard approach for non-uniform bidding in FPA~\cite{karlsson2021adaptive,pan2020bid,qu2024double,slikker2007bidding,fagandini2023computing,zhou2021efficient}. Mainstream methods fall into two categories. The first~\cite{karlsson2021adaptive,fagandini2023computing} predicts an optimal shading factor, $bid_{j}=shading\_factor_j*\eta*Value_j$. Due to estimation variance, even unbiased predictors can underbid and lose many auctions; asymmetric losses can penalize underbidding but provide no guarantee~\cite{gligorijevic2020bid}. The other~\cite{ren2019deep,pan2020bid,zhou2021efficient} models the win probability and searches the bid that maximizes expected surplus, $bid_{j}= \arg\max_{b\ge 0}(\eta*Value_j-b)p_j(b) $. Related advances include leveraging second-price signals to help first-price modeling~\cite{huang2024second} and end-to-end multi-slot bid shading~\cite{gong2023mebs}. Crucially, these methods are all designed for a pure FPA environment and do not account for organic-paid coexistence.
\section{Algorithms}
For convenience, we list the notation used throughout the paper.

\label{Alogrithms}
\noindent\begin{tabularx}{\linewidth}{l@{\hspace{1em}}L}
\hline
$\eta_i$                    & control parameter of bidding strategy in channel $i$ \\
$v_{ij}(\eta_i)$              & value of impression $j$ in channel $i$  \\
$b_{ij}(\eta_i)$              & bid price of impression $j$ in channel $i$  \\
$wp_{ij}$              & winning price of impression $j$ in channel $i$ \\
$p_{ij}(b)$          & win probability of impression $j$ in channel $i$ at bid $b$\\
$V_i(\eta_i)$              & total value in channel $i$ \\
$C_i(\eta_i)$              & total cost in channel $i$ \\
$MC_i(\eta_i)$                & marginal cost in channel $i$ \\
\hline
\end{tabularx}

\subsection{Problem Formulation}
\label{sec:Problem_Formu}
We consider a unified optimization problem over heterogeneous channels $\mathcal{I} = \{su, fu, fn\}$ corresponding to SPA, uniform FPA, and non-uniform FPA, respectively. The objective is defined under either MaxReturn ($B > 0, R = 0$) or TargetROAS ($B > 0, R > 0$) objectives, subject to Budget and TargetROAS constraints:
\begin{equation}
\label{eq:opt}
\max_{\{\eta_i\}} \sum_{i \in \mathcal{I}} V_i(\eta_i) \quad \text{s.t. } \sum_{i \in \mathcal{I}} C_i(\eta_i) \le B, \quad \textstyle\frac{\sum_i V_i(\eta_i)}{\sum_i C_i(\eta_i)} \ge R.
\end{equation}
Assuming $V_i, C_i$ are continuously differentiable, the optimal solution can be characterized via the KKT conditions \citep{tabak1971optimal}.

\begin{theorem}[Marginal Cost Equalization]
\label{thm:marginal_cost_equalization}
For heterogeneous channels sharing a common \text{Budget} B and \text{TargetROAS} R, any regular interior optimum satisfies the MC equalization condition:
\begin{equation}
\mathrm{Marginal Cost}_{i}(\eta_i^*)\ =\frac{C_i^{'}(\eta_i^{*})}{V_i^{'}(\eta_i^{*})}
=\mu, \forall i\in\{su,fu,fn\}
\end{equation}
where $\mu$ is a channel-independent constant.
\end{theorem}

\begin{proof}[Proof Sketch]
Let $L(\eta, \lambda, \gamma) = \sum V_i - \lambda(R \sum C_i - \sum V_i) - \gamma(\sum C_i - B)$ be the Lagrangian. The stationarity condition $\partial L / \partial \eta_i = 0$ implies $(1+\lambda)V_i'(\eta_i^*) = (\lambda R + \gamma)C_i'(\eta_i^*)$. Rearranging terms gives $C_i'/V_i' = (1+\lambda)/(\lambda R + \gamma) \triangleq \mu$. The full proof is provided in the Appendix \ref{mc_eq_proof}.
\end{proof}

% \begin{theorem}[Marginal Cost Equalization]
% \label{thm:mc}
% At optimality, all channels share a constant marginal cost $\mu$:
% \begin{equation}
% MC_i(\eta_i^*) \triangleq \textstyle\frac{C_i'(\eta_i^*)}{V_i'(\eta_i^*)} = \mu, \quad \forall i \in \mathcal{I}.
% \end{equation}
% \end{theorem}

% \begin{proof}[Proof sketch]
% Let $L(\eta, \lambda, \gamma) = \sum V_i - \lambda(R \sum C_i - \sum V_i) - \gamma(\sum C_i - B)$ be the Lagrangian. The stationarity condition $\partial L / \partial \eta_i = 0$ implies $(1+\lambda)V_i'(\eta_i^*) = (\lambda R + \gamma)C_i'(\eta_i^*)$. Rearranging terms gives $C_i'/V_i' = (1+\lambda)/(\lambda R + \gamma) \triangleq \mu$.
% \end{proof}

\begin{figure*}[t!]
  \centering
  \includegraphics[width=0.85\linewidth]{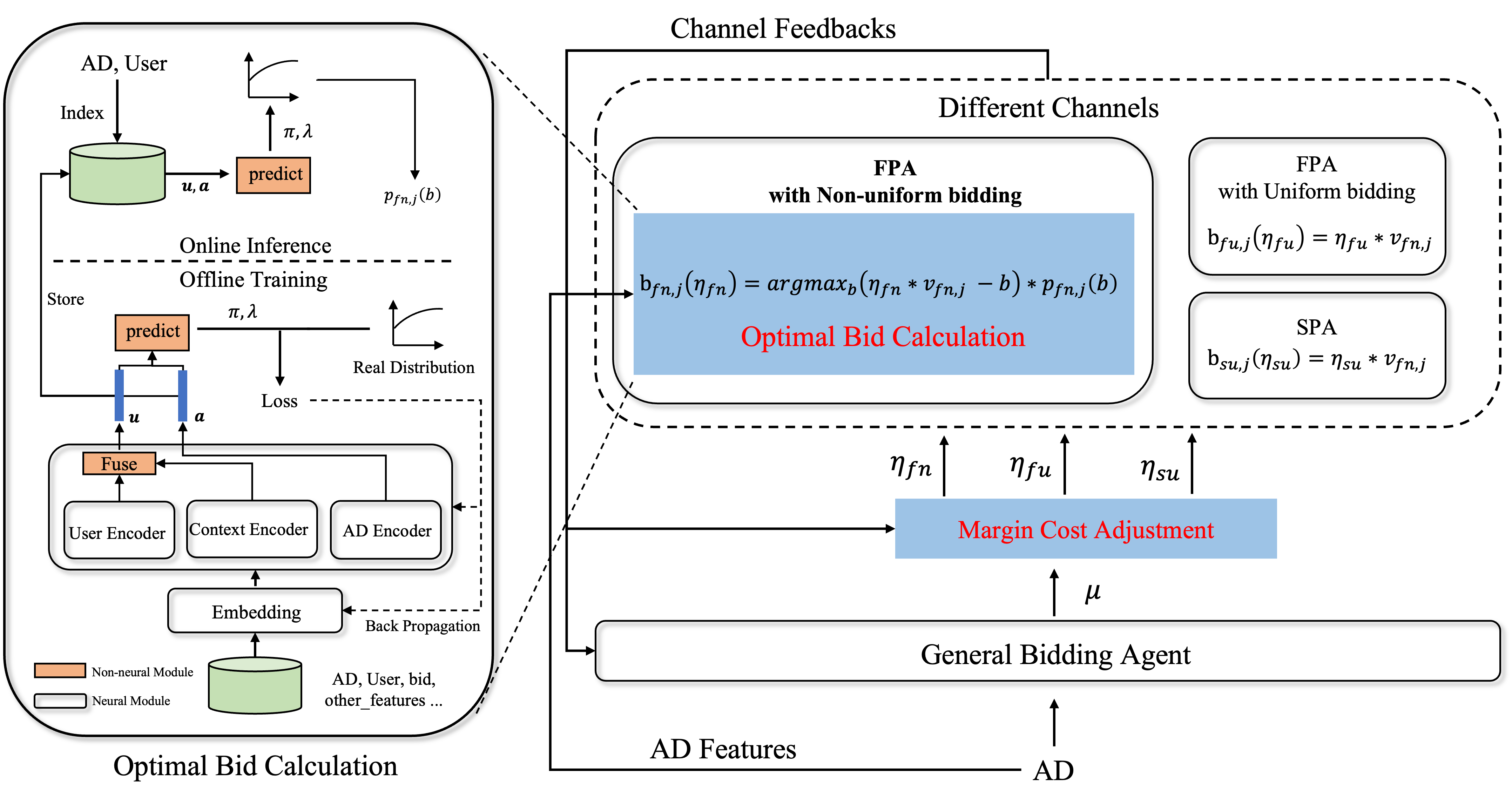}
  \caption{Overview of HOB. The left module performs request-level bid optimization for FPA-NU by modeling free-win probability and winning-price uncertainty. The right module aligns channel-level MC through a shared target, enabling globally coordinated autobidding across heterogeneous channels.}
  \Description{a near-real-time control loop.}
  \label{Fig:architecture}
\end{figure*}

\textbf{Our basic idea} is to optimize bidding holistically in heterogeneous bidding environments by using channel-specific bidding and aligning MC across channels. Specifically, Sections 3.2–3.4 derive the bidding formulation and MC for each channel type, while Section 3.5 shows how these channel-specific MCs can be aligned through a shared target for global coordination (see Figure~\ref{Fig:architecture}).

\subsection{Bid and MC in SPA}
\label{Optimal Bid SPA}
Under the Generalized Second-Price (GSP) auction rule, a common variant of the SPA, the winner pays the minimum amount required to maintain their rank over the next-highest bidder. Uniform bidding is optimal in this setting, so we do not consider non-uniform bidding. The cost of acquiring incremental impressions is proportional to the bid multiplier $\eta_{su}$, and the MC equals $\eta_{su}$: 

\begin{equation}
\begin{aligned}
V_{su}(\eta_{su}) &= \sum_{j\in\mathcal J_{su}} v_{su,j}\cdot\mathbb I\!\{b_{su,j}(\eta_{su})\ge wp_{su,j}\},\\
C_{su}(\eta_{su}) &= \sum_{j\in\mathcal J_{su}} wp_{su,j}\cdot\mathbb I\!\{b_{su,j}(\eta_{su})\ge wp_{su,j}\},\\
b_{su,j}(\eta_{su})&= v_{su,j}\cdot\eta_{su},\qquad
MC_{su}(\eta_{su})=\eta_{su}.
\label{mc_spa}
\end{aligned}
\end{equation}

% \noindent\textbf{FPA channel with uniform bidding.}
\subsection{Bid and MC in uniform FPA}
In FPA channels, the winner's payment is equal to their submitted bid. When increasing their bid multiplier from $\eta_{fu}$ to $\eta_{fu}+\Delta$, they not only pay for newly won impressions but also pay an additional $\Delta \cdot V(\eta_{fu})$ for the impressions they were already won. This additional term results in a higher MC:
\begin{equation}
\begin{aligned}
V_{fu}(\eta_{fu}) &= \sum_{j\in\mathcal J_{fu}} v_{fu,j}\cdot\mathbb I\!\{b_{fu,j}(\eta_{fu})\ge wp_{fu,j}\},\\
C_{fu}(\eta_{fu}) &= \sum_{j\in\mathcal J_{fu}} b_{fu,j}(\eta_{fu})\cdot\mathbb I\!\{b_{fu,j}(\eta_{fu})\ge wp_{fu,j}\},\\
b_{fu,j}(\eta_{fu})&= v_{fu,j}\cdot\eta_{fu},\qquad
MC_{fu}(\eta_{fu})=\eta_{fu}+\frac{V(\eta_{fu})}{V'(\eta_{fu})}.
\label{eq:mc_fpa_ub}
\end{aligned}
\end{equation}
The full derivation of $MC_{fu}$ can be found in Appendix \ref{mc_fu_proof}.
% The winner pays its bid. Increasing the $\eta$ raises payments for both newly and previously won impressions, resulting in higher marginal cost:
% \begin{equation}
% \begin{aligned}
% % &V_{fu}(\eta_{fu})=\mathbb E\!\left[v_{fu,j}\cdot\mathbb I\{b_{fu,j}\ge wp\}\right],
% % b_{fu,j}(\eta_{fu})= v_{fu,j}\cdot\eta_{fu},\\
% % &C_{fu}(\eta_{fu})=\mathbb E\!\left[b_{fu,j}\cdot\mathbb I\{b_{fu,j}\ge wp\}\right],
% % MC_{fu}(\eta_{fu})=\eta_{fu}+\frac{V(\eta_{fu})}{V'(\eta_{fu})}.
% &b_{fu,j}(\eta_{fu})= v_{fu,j}\cdot\eta_{fu},
% MC_{fu}(\eta_{fu})=\eta_{fu}+\textstyle\frac{V(\eta_{fu})}{V'(\eta_{fu})}.
% \label{eq:mc_fpa_ub}
% \end{aligned}
% \end{equation}

% \begin{proof}[Proof sketch]
% The expected cost satisfies
% $C_{fu}(\eta)=\eta\,V_{fu}(\eta)$.
% Differentiating gives $C_{fu}'(\eta)=V_{fu}(\eta)+\eta V_{fu}'(\eta)$.
% By definition, the marginal cost is
% $
% MC_{fu}(\eta)= \textstyle \frac{dC_{fu}/d\eta}{dV_{fu}/d\eta}
% = \textstyle \frac{C_{fu}'(\eta)}{V_{fu}'(\eta)}
% =\eta+ \textstyle \frac{V_{fu}(\eta)}{V_{fu}'(\eta)}.
% $
% \end{proof}

\subsection{Bid and MC in non-uniform FPA}
\label{Optimal Bid FPA nub}
In FPA channels that allow non-uniform bidding, as discussed in the related work, there are two typical forms of non-uniform bidding. Compared with shading-factor methods that directly scale bids but do not provide an explicit expression of MC, surplus maximization with winning probability $p_{fn,j}(b)$ yields an explicit MC. The optimal bid maximizes the $\eta_{fn}$-weighted expected surplus~\cite{karlsson2021adaptive}:
\begin{equation}
b_{fn,j}(\eta_{fn})
=\arg\max_{b\ge 0}\; \bigl(\eta_{fn}\, v_{fn,j} - b\bigr)\,p_{fn,j}(b).
\label{eq:surplus_continuous}
\end{equation}

Here we work with ex-ante expected value and cost rather than replay-style indicator quantities. In non-uniform FPA, the winning price is modeled as a random variable, and the bid optimization in Eq.~\eqref{eq:surplus_continuous} is defined against its win probability $p_{fn,j}(b)$. Consequently, the aggregate value and cost are naturally defined as expectations over the winning-price distribution:
\begin{equation}
\begin{aligned}
V_{fn}(\eta_{fn}) = \sum_{j\in J_{fn}} v_{fn,j}\, p_{fn,j}(b_{fn,j}(\eta_{fn})),\\
C_{fn}(\eta_{fn}) = \sum_{j\in J_{fn}} b_{fn,j}(\eta_{fn})\, p_{fn,j}(b_{fn,j}(\eta_{fn})).
\end{aligned}
\end{equation}

\begin{theorem}[MC in non-uniform FPA Channel]
\label{thm:marginal_cost}
For the optimal bid in Eq.~\eqref{eq:surplus_continuous}, the MC is equal to the control parameter $\eta_{fn}$.
\end{theorem}

\label{mc_fn_proof_sketch}
\begin{proof}[Proof Sketch]
For each request $j$ (omitting the subscript $fn$ for brevity), the FOC yields 
$p_j(b_j) + b_j p'_j(b_j) = \eta v_j p'_j(b_j)$. 
Defining aggregate value $V(\eta) = \sum_j v_j p_j(b_j)$ and cost $C(\eta) = \sum_j b_j p_j(b_j)$, their derivatives satisfy:
\begin{equation*}
C'(\eta) = \sum_j [p_j + b_j p'_j] b'_j(\eta) = \sum_j (\eta v_j p'_j) b'_j(\eta) = \eta V'(\eta).
\end{equation*}
Thus, $MC_{fn}(\eta_{fn}) = \textstyle \frac{dC/d\eta_{fn}}{dV/d\eta_{fn}} = \eta_{fn}$. The full proof is provided in Appendix \ref{mc_fn_proof}.
\end{proof}
\paragraph{Remark 3.1 (Boundary cases).}
In practice, we constrain the bid to the feasible interval $[0,\eta_{fn} v_{fn,j}]$ and project the unconstrained solution onto this range when necessary. Thus, $MC_{fn}(\eta_{fn})=\eta_{fn}$ should be understood for the interior active set, while the projected solution is used in deployment.

\noindent\textbf{Winning-Price Modeling.} We model the $p_{fn,j}(b)$ using a Zero-Inflated Exponential (ZIE) distribution:
$p_{fn,j}(b| u, a) = \pi_{\theta} + (1-\pi_{\theta}) \cdot (1-e^{-\lambda_{\theta} \cdot b})$, where $\pi_{\theta}$ and $\lambda_{\theta}$ are predicted by a neural model conditioned on user $u$ and advertiser $a$. The ZIE model is chosen for its: (i) High Fidelity: it captures the zero-spike (due to high UE scores) and right-skewed tail of competitive auctions; (ii) Robustness: its parsimonious two-parameter form prevents overfitting; and (iii) Efficiency: it facilitates fast optimal bid computation. The surplus maximizer admits a unique closed-form solution:
$
    b_{fn,j}^* = \eta_{fn} v_{fn,j} + \textstyle\frac{1 - \omega\bigl( 1 + \lambda_{\theta} \eta_{fn} v_{fn,j} - \ln(1 - \pi_{\theta}) \bigr)}{\lambda_{\theta}}
    % \label{eq:wright_solution}
    $
, where $\omega(\cdot)$ is the Wright $\omega$ function~\cite{corless2002wright}. The strictly increasing nature of $\omega(z)$ guarantees a unique optimum in $[0, \eta_{fn} v_{fn,j}]$, allowing for efficient online inference via mature numerical libraries.

\subsection{MC Alignment}
\label{sec:MCA}
% Without loss of generality, we assume $\eta_{su}=\eta_{fn}=\mu$. Then the only thing we should do is to let $\eta_{fu} + \frac{V(\eta_{fu})}{V'(\eta_{fu})} = \mu$. The term $V(\eta_{fu}) / V'(\eta_{fu})$ required for this alignment can be computed via intensive impressions replay. However, for practical implementation, we adopt a simpler approximation. Since the value function is expected to be monotonically increasing with diminishing marginal returns (i.e., $V'(\eta_{fu})>0$ and $V''(\eta_{fu})<0$), we approximate it with a shifted power-law form:

To achieve global optimality, we align all channels to a shared MC $\mu$. Since $MC_{su} = \eta_{su}$ and $MC_{fn} = \eta_{fn}$, this directly yields $\eta_{su} = \eta_{fn} = \mu$. For uniform FPA, since $MC_{fu} = \eta_{fu} + V_{fu}(\eta_{fu})/V'_{fu}(\eta_{fu}) = \mu$. we need a non-trivial mapping from $\mu$ back to $\eta_{fu}$. The term $V_{fu}(\eta_{fu}) / V'_{fu}(\eta_{fu})$ required for this alignment. For practical implementation, we adopt a simpler approximation. Since the value function is expected to be monotonically increasing with diminishing marginal returns (i.e., $V'_{fu}(\eta_{fu})>0$ and $V''_{fu}(\eta_{fu})<0$), we approximate it with a shifted power-law form: $a \cdot (\eta_{fu}+c)^b$ ($a,c>0, 0<b<1$). By computing the ratio $V_{fu}/V'_{fu}$ and substituting it into the equation, we obtain a closed-form alignment mapping:
$ \eta_{fu} = \textstyle\frac{b \cdot \mu - c}{b+1}$.

\begin{theorem}[Optimality of $\mu^*$]
\label{thm:exist_uni_mu}
At the aggregate level, we adopt the mild assumption that $V(\mu)$ and $C(\mu)$ are non-decreasing in $\mu$, and that $\mathrm{ROAS}(\mu)=V(\mu)/C(\mu)$ is non-increasing on the feasible region. Consider
\[
\max_\mu V(\mu)\quad \text{s.t.}\quad C(\mu)\le B,\ \mathrm{ROAS}(\mu)\ge R.
\]
Define $\mu_B^*=\sup\{\mu\mid C(\mu)\le B\}, \mu_R^*=\sup\{\mu\mid \mathrm{ROAS}(\mu)\ge R\}.$
% Given the monotonic properties where $V(\mu)$, $C(\mu)$ are non-decreasing and $ROAS(\mu) = V(\mu)/C(\mu)$ is non-increasing under the stated assumptions.
% \begin{equation*}
%     \max_\mu V(\mu) \quad \text{s.t.} \quad C(\mu) \le B, \quad ROAS(\mu) \ge R,
% \end{equation*}

% Define $\mu_B^* = \sup\{\mu \mid C(\mu) \le B\}$ and $\mu_R^* = \sup\{\mu \mid ROAS(\mu) \ge R\}$.
Then an optimal solution is given by $\mu^* = \min(\mu_B^*, \mu_R^*)$. Moreover, $\mu^*$ is the largest feasible aligned threshold. If the feasible frontier is strictly monotone, then this optimal solution is unique.

\end{theorem}
\begin{proof}[Proof Sketch]
Since $V(\mu)$ is non-decreasing, maximizing $V(\mu)$ is equivalent to choosing the largest $\mu$ that satisfies both constraints. By the monotonicity of $C(\mu)$ and $\mathrm{ROAS}(\mu)$, the feasible set is characterized by $\mu \le \mu_B^* \quad \text{and} \quad \mu \le \mu_R^*.$
Hence the largest feasible threshold is $\mu^*=\min(\mu_B^*,\,\mu_R^*)$
which is therefore optimal. Uniqueness further follows when the feasible frontier is strictly monotone. The full proof is provided in Appendix~\ref{opt_u_proof}.

% Since $V(\mu)$ is non-decreasing, the optimal $\mu$ is the largest value that simultaneously satisfies both $C(\mu) \le B$ and $ROAS(\mu) \ge R$. Given the monotonicity of Cost and ROAS, this is uniquely given by $\min(\mu_B^*, \mu_R^*)$. The full proof is provided in the Appendix \ref{opt_u_proof}.
\end{proof}

% In practice, because the aggregate  curves are only indirectly observed and may drift over time, we maintain a time-varying threshold $\mu_t$ and update it online via PID control to track the largest feasible operating point under the budget and ROAS constraints.
% Appendix~\ref{appendix:Pseudo-code} provides the pseudo-code.

% Theorem 3.3 characterizes the optimal static aligned threshold $\mu^*$. In practice, rather than solving $\mu^*$ explicitly from aggregate curves, we maintain a dynamic threshold $\mu_t$ and update it online via PID control using budget/ROAS feedback, thereby steering the system toward the largest feasible aligned operating point.
Theorem~\ref{thm:exist_uni_mu} defines the optimal static threshold $\mu^*$. In practice, since future traffic is unknown, $\mu^*$ cannot be solved in advance. We therefore maintain a time-varying threshold $\mu_t$ and update it online from budget/ROAS feedback, e.g., via PID control. Shifted power-law coefficients are updated via nonlinear least squares on a rolling 24-hour log window. Appendix~\ref{appendix:Pseudo-code} provides the pseudo-code.

\begin{table}[ht]
\centering
\scriptsize
\setlength{\tabcolsep}{4pt} % 稍微增加点间距让其填满
\caption{YOYI experimental results: Metrics per channel}
\label{tab:metrics_per_channel}
    \begin{tabular*}{\linewidth}{@{\extracolsep{\fill}} ll rrr @{}}
    \toprule
    \textbf{Method} & \textbf{Channel} & \textbf{Pred. Clk}$\uparrow$ & \textbf{Cost} & \textbf{MC} \\
    \midrule
    \multirow{3}{*}{UB}   & FPA-u  & 6307 & 4005 & 1.35 \\
                          & FPA-nu & 6307 & 4005 & 1.35 \\
                          & SPA-u  & 6307 & 1988 & 0.64 \\
                          & All    & 18921 & 9999.8 & Std(MC)=0.34 \\
    \midrule
    \multirow{3}{*}{NUB-Z}  & FPA-u  & 7600 & 5902 & 2.02 \\
                          & FPA-nu & 3333 & 1134 & 0.55 \\
                          & SPA-u  & 7600 & 2905 & 0.78 \\
                          & All & 18533 & 9942.5 & Std(MC)=0.64 \\
    \midrule
    \multirow{4}{*}{\makecell[l]{\textbf{HOB-Z}}} & FPA-u & 5417 & 2849 & {1.07} \\
                          & FPA-nu & 5515 & 2889 & {0.95} \\
                          & SPA-u  & 9158 & 4260 & {0.99} \\
                          & \textbf{All} & \textbf{20090} & \textbf{9999} & \textbf{Std(MC)=0.05} \\
    \bottomrule
    \end{tabular*}
\end{table}

% \begin{table}[ht]
% \centering
% \scriptsize
% \setlength{\tabcolsep}{4pt}
% \caption{YOYI experimental results: Ablation study}
% \label{tab:distribution_ablation}
% \renewcommand{\arraystretch}{1.24}
%     \begin{tabular*}{\linewidth}{@{\extracolsep{\fill}} l cc @{}}
%     \toprule
%     \multirow{2}{*}{\textbf{Method}} & \textbf{MaxReturn} & \textbf{TargetCPC} \\
%     \cmidrule{2-3}
%     & \textbf{Pred. Clk$\uparrow$} & \textbf{Pred. Clk$\uparrow$} \\
%     \midrule
%     UB     & +0.0\% & +0.0\% \\
%     \midrule
%     NUB-G     & -7.5\% & -9.0\% \\
%     NUB-L     & -9.4\% & -12.4\% \\
%     NUB-Z     & -2.8\% & -3.4\% \\
%     \midrule
%     HOB-G   & +2.3\% & +2.9\% \\
%     HOB-L   & +2.5\% & +3.2\% \\
%     \textbf{HOB-Z}     & \textbf{+4.8\%} & \textbf{+6.0\%} \\
%     HOB-Z-oracle     & +6.3\% & +7.6\% \\
%     \bottomrule
%     \end{tabular*}
% \end{table}

\begin{table}[t]
\centering
\caption{YOYI experimental results: Ablation study.% Results are reported as mean $\pm$ std over 10 independent runs.
 $\dagger$ denotes our method significantly outperforms the baselines with $p < 0.05$ under a paired t-test over 10 independent runs.
}
\label{tab:distribution_ablation}
\begin{tabular}{@{} l @{\hspace{18pt}} c @{\hspace{18pt}} c @{}}
\toprule
\multirow{2}{*}{\textbf{Method}} & \textbf{MaxReturn} & \textbf{TargetCPC} \\
 & Pred.\ Clk$\uparrow$ & Pred.\ Clk$\uparrow$ \\
\midrule
UB          & \phantom{**}+0.0\%\phantom{{\scriptsize$\pm$0.0}*} & \phantom{**}+0.0\%\phantom{{\scriptsize$\pm$0.0}*} \\
\midrule
NUB-G       & $-$7.5\%\phantom{**}  & $-$9.0\%\phantom{**} \\
NUB-L       & $-$9.4\%\phantom{**}  & $-$12.4\%\phantom{**} \\
NUB-Z       & $-$2.8\%\phantom{**}  & $-$3.4\%\phantom{**} \\
\midrule
HOB-G       & +2.3\%\phantom{*}   & +2.9\%\phantom{*} \\
HOB-L       & +2.5\%\phantom{*}   & +3.2\%\phantom{*} \\
\textbf{HOB-Z} & \textbf{+4.8}\%$^{\dagger}$ & \textbf{+6.0}\%$^{\dagger}$ \\
HOB-Z-oracle & +6.3\%\phantom{*} & +7.6\%\phantom{*} \\
\bottomrule
\end{tabular}

\end{table}

% \begin{table}[!t]
%   \centering
%   \captionsetup{skip=6pt}
%   \caption{Correlation: Model Fit and Gains.}
%   \label{tab:final_comparison_correct}
%   \begin{tabular}{@{}lcc@{}} % l=left, c=center
%     \toprule
%     Model& BCE & Surplus Rate    \\
%     \midrule
%     Exponential  & 0.96 & 54.15\%   \\
%     Log-Normal  & 0.61 & 79.89\%   \\
%     Gamma & 0.57 & 81.29\%   \\
%     \textbf{ZIE}    & \textbf{0.54} & \textbf{83.14\%}  \\
%     \bottomrule
%   \end{tabular}
% \end{table}

\section{Experiments}\label{exp}
In our evaluation, we systematically investigate four key Research Questions (RQs):
(1) Does MCA improve performance?
(2) What are the benefits of ZIE distribution and power-law approximation?
(3) Is HOB robust to different settings?
(4) How can HOB be deployed in industrial systems for efficiency gains?

Because no public dataset contains heterogeneous auction channels with organic-paid coexistence, we construct a controlled benchmark from YOYI for reproducible offline comparison, while relying on production logs and online A/B tests for realism.

\subsection{Offline controlled benchmark}
\textbf{YOYI} \cite{user_response_learning} contains 402M impressions, 500K clicks and 428K CNY expense, among which 363M impressions are used for training and 39M for testing. Each impression is represented as a tuple $(y,z,\mathbf{x})$, where $y \in \{0,1\}$ denotes whether the ad was clicked, $z$ denotes the winning price, and $\mathbf{x}$ is a feature vector describing the request. Since YOYI doesn't provide the predicted value for each impression, we assign each impression an extra predicted value $v$ using a neural model, which is trained by treating $y$ as label. To further simulate the heterogeneous channels, the dataset is split equally into three parts, where distinct auction mechanisms are adopted. Specifically, we consider three different types of channel: SPA channel (SPA-u), FPA channel restricted to uniform bidding (FPA-u), FPA channel permitting non-uniform bidding (FPA-nu). We assess performance using these key metrics: Predicted Click, Cost, MC.

\noindent \textbf{To answer RQ(1):} We compare the following three strategies:
\begin{itemize}
    \item \textbf{UB:} Unified $\eta$ and uniform bidding for all three channels.
    \item \textbf{NUB:} Unified $\eta$; SPA and FPA-u use uniform bidding, while FPA-nu uses the non-uniform bid in Section~\ref{Optimal Bid FPA nub}.
    \item\textbf{HOB(MCA\&NUB):} Aligns marginal costs ($MC=\mu$) via channel-specific $\eta_i$ and non-uniform bidding for FPA-nu.
\end{itemize}

% As shown in Table~\ref{tab:metrics_per_channel}, \textbf{HOB} achieves superior overall performance by aligning MC across channels.
Table \ref{tab:metrics_per_channel} reports the MCA evaluation results on the YOYI dataset. Two observations stand out. First, bid shading without cross-channel MC alignment degrades performance, confirming that local optimization alone is insufficient. Second, HOB achieves the best overall performance by driving substantially better cross-channel alignment, reducing the standard deviation of channel-wise MC to 0.05.

\begin{figure}[!t]
  \centering
  \includegraphics[width=\linewidth]{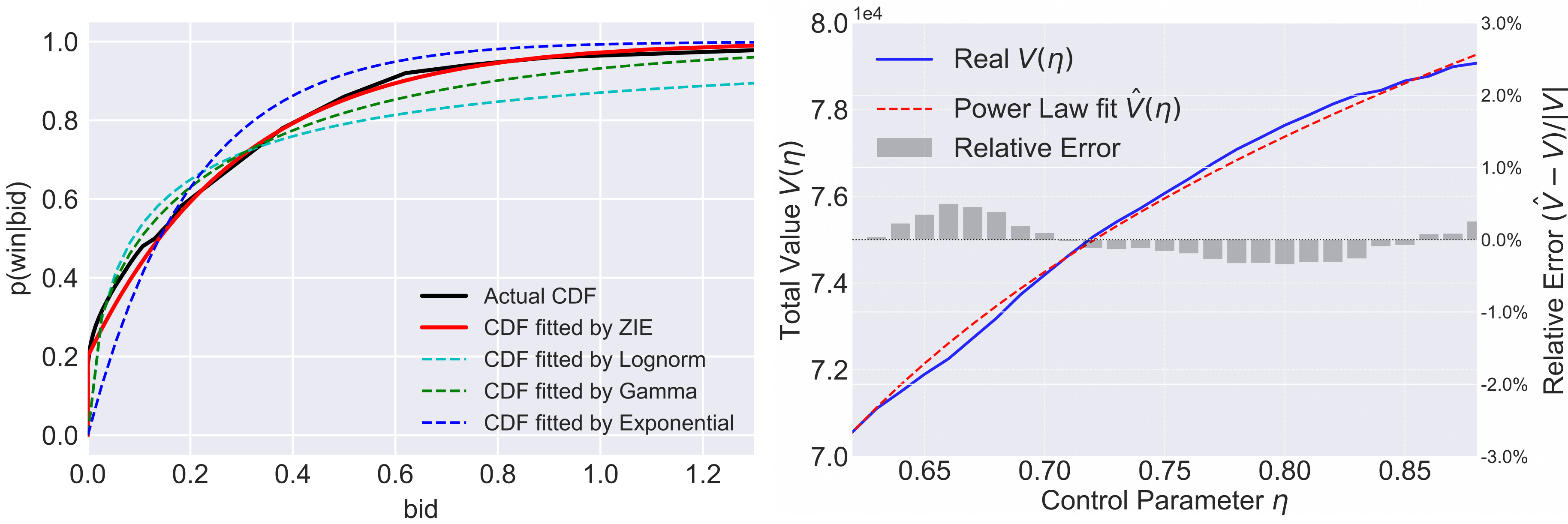}
  \caption{(a) Empirical CDF (black) vs.\ fitted CDFs under different distributions. (b) Validation of Power-law assumption.}
  \Description{MC_fig}
    \label{fig:power-law-validation}
\end{figure}

\noindent
\textbf{To answer RQ(2):} We examine the validity of the ZIE distribution and the power-law approximation.

\textbf{Winning-Price Fitting.}
Since the YOYI dataset contains no free wins, it cannot directly validate HOB's handling of zero-price impressions. To introduce a realistic zero-price component, we inject zero-price impressions by adding zero-mean Gaussian noise ($\sigma=1.0z$) to winning prices and clipping at zero, yielding 15.41\% zero-price impressions. We then estimate $p(b)$ with DeepFM as it has been verified to be highly effective in ~\cite{zhou2021efficient}, to predict these distribution parameters from $\mathbf{x}$ by MLE, and evaluate Gamma (G), Log-normal (L), and ZIE (Z) under \textbf{NUB} and \textbf{HOB}. Table~\ref{tab:distribution_ablation} shows that HOB-Z consistently performs best. To further validate these assumptions in practice, we fit each candidate distribution to production bidding logs. As shown in Figure~\ref{fig:power-law-validation}(a), the ZIE distribution most accurately captures the empirical CDF.

\textbf{Value-Function Approximation.}
Replacing the power-law-based $\eta_{fu}$ with a log-replay oracle (HOB-Z-oracle) yields only a \textbf{1.5\%} Clk gain on average in Table~\ref{tab:distribution_ablation}, indicating the approximation is sufficiently accurate. Figure~\ref{fig:power-law-validation}(b) further shows that the fitted curve closely matches the ground-truth value curve.

\begin{figure}[!t]
  \centering
  \includegraphics[width=\linewidth]{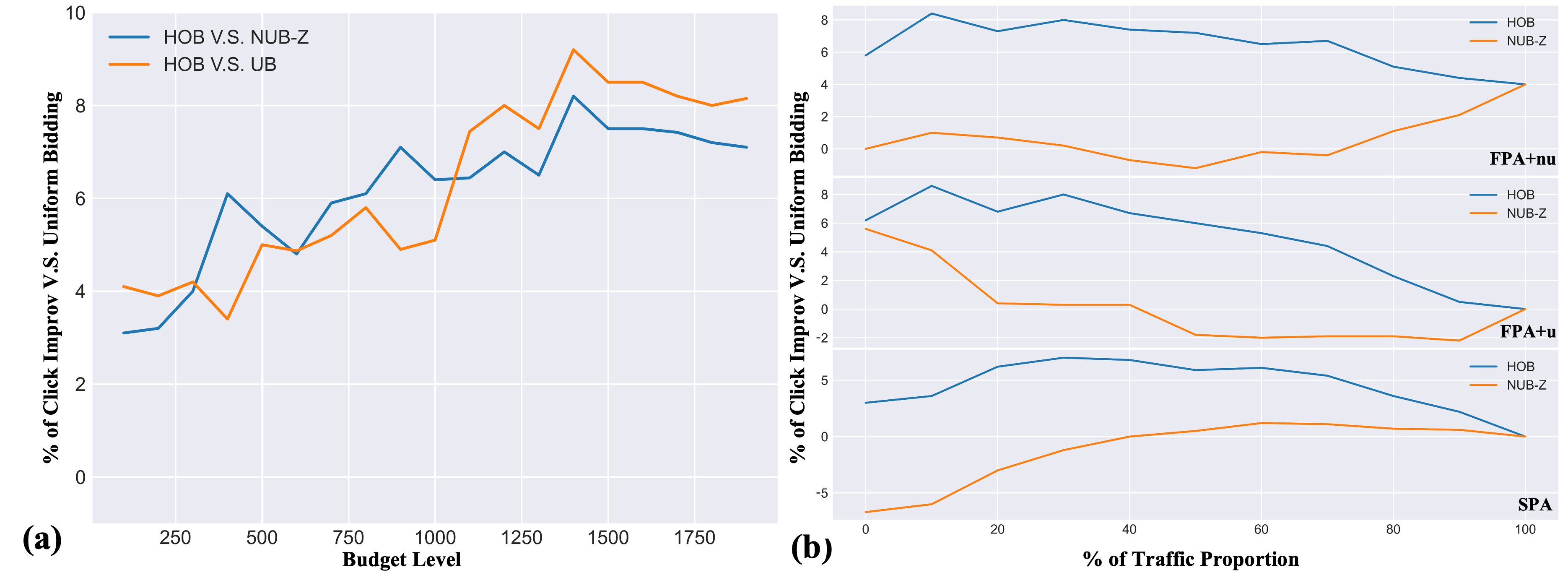}
  \caption{(a) HOB gains under different budget levels. (b) HOB gains under different channel proportions.}
  \Description{MC_fig}
    \label{fig:budget-channel-comp}
\end{figure}

% \subsection{Robustness to Property Variations}
\noindent
\textbf{To answer RQ(3):} We test robustness under different settings. From the advertiser perspective, Figure~\ref{fig:budget-channel-comp} (a) shows the gains in Click under varying budget levels, compared with \textbf{UB} and \textbf{NUB}. \textbf{HOB} delivers consistent gains across budget levels, with larger improvements for high-spend campaigns due to greater savings on free-win impressions. We also vary the composition of heterogeneous channels. Keeping the total number of impressions fixed, we increase the share of one channel and split the remaining evenly between the other two. As shown in Figure~\ref{fig:budget-channel-comp} (b), \textbf{NUB-Z} fails to consistently outperform \textbf{UB}, while \textbf{HOB} demonstrates robust superiority across all channel-share settings, with the largest gains under balanced traffic mixes (i.e., no single channel dominates).

% \begin{figure}[h]
%   \centering
%   \includegraphics[width=\linewidth]{figures/budget_comp.png}
%   \caption{HOB's improvements on different budget levels.}
%   \Description{MC_fig}
%     \label{fig:budget-comp}
% \end{figure}

% \begin{figure}[h]
%   \centering
%   \includegraphics[width=\linewidth]{figures/channel_comp.png}
%   \caption{Analysis on different channel proportions.}
%     \label{fig:channel-comp}
% \end{figure}

\subsection{Offline Industrial log replay}
In addition, we conduct offline evaluation on a 0.1\% sample of production auction logs, containing tens of millions of requests from MaxReturn and TargetROAS campaigns. Since true counterfactual conversions are unobservable, we replay each method using the platform's production value model to assign predicted values, and compare methods on aggregated predicted GMV, cost, and ROAS. Notably, 60\% of the GMV in this dataset originates from a non-uniform FPA channel. The dataset provides rich features for users (e.g., gender, age, purchase history) and ads (e.g., shop, brand, category) enabling the training of winning price models under diverse assumptions. 7-day training set and 1-day test are set based on a temporal split. Table~\ref{tab:exp_Industrial} reports the results for both MaxReturn and TargetROAS campaigns. Under the TargetROAS objective, higher spend is desirable as long as the ROAS constraint remains satisfied. The increased spend of HOB therefore indicates that it unlocks additional efficient traffic rather than relaxing the constraint.

% \begin{table}[ht]
% \centering
% \scriptsize
% \setlength{\tabcolsep}{4pt}
% \caption{Industrial log-replay results on the production dataset. Budget = 300.00 for MaxReturn, and TargetROAS = 3.00 for TargetROAS.}
% \label{tab:exp_Industrial}
% \renewcommand{\arraystretch}{1.24}
% \begin{tabular*}{\linewidth}{@{\extracolsep{\fill}} l ccc ccc @{}}
% \toprule
% \multirow{2}{*}{\textbf{Method}} 
% & \multicolumn{3}{c}{\textbf{MaxReturn}} 
% & \multicolumn{3}{c}{\textbf{TargetROAS}} \\
% \cmidrule(lr){2-4} \cmidrule(lr){5-7}
% & \textbf{Pred. GMV$\uparrow$} & \textbf{Cost} & \textbf{ROAS$\uparrow$} 
% & \textbf{Pred. GMV$\uparrow$} & \textbf{Cost} & \textbf{ROAS} \\
% \midrule
% UB     
% & 907.1 (+0.0\%) & 300.0 & 3.0 
% & 909.6 (+0.0\%) & 303.2 & 3.0 \\
% \midrule
% NUB-G  
% & 910.6 (+0.4\%) & 300.0 & 3.1 
% & 922.8 (+1.5\%) & 308.1 & 3.0 \\
% NUB-L  
% & 920.2 (+1.5\%) & 299.9 & 3.1 
% & 931.6 (+2.4\%) & 310.6 & 3.0 \\
% NUB-Z  
% & 909.6 (+0.3\%) & 300.00 & 3.0  
% & 920.5 (+1.2\%) & 306.8 & 3.0 \\
% \midrule
% MCA\&NUB-G
% & 1011.9 (+11.6\%) & 299.9 & 3.4 
% & 1053.7 (+15.8\%) & 351.4 & 3.0 \\
% MCA\&NUB-L 
% & 1014.0 (+11.8\%) & 299.9 & 3.4 
% & 1054.5 (+15.9\%) & 351.5 & 3.0 \\
% \textbf{HOB-Z} 
% & \textbf{1024.0 (+12.9\%)} & 300.0 & 3.4
% & \textbf{1065.4 (+17.1\%)} & 355.1 & 3.0 \\
% \bottomrule
% \end{tabular*}
% \end{table}

\begin{table}[t]
\centering
\caption{Industrial log-replay results. Budget = 300.00 for MaxReturn, and TargetROAS = 3.00 for TargetROAS. $\dagger$ denotes our method significantly outperforms the baselines with $p < 0.05$ under a paired t-test over 10 independent runs.
}
\label{tab:exp_Industrial}
\resizebox{\columnwidth}{!}{%
\begin{tabular}{l ccc ccc}
\toprule
\multirow{2}{*}{\textbf{Method}} & \multicolumn{3}{c}{\textbf{MaxReturn}} & \multicolumn{3}{c}{\textbf{TargetROAS}} \\
\cmidrule(lr){2-4} \cmidrule(lr){5-7}
 & Pred.\ GMV$\uparrow$ & Cost & ROAS$\uparrow$ & Pred.\ GMV$\uparrow$ & Cost & ROAS \\
\midrule
UB 
 & 907.1\phantom{*} (+0.0\%) 
 & 300.0 & 3.0 
 & 909.6\phantom{*} (+0.0\%) 
 & 303.2 & 3.0 \\
\midrule
NUB-G 
 & 910.6\phantom{*} (+0.4\%) 
 & 300.0 & 3.1 
 & 922.8\phantom{*} (+1.5\%) 
 & 308.1 & 3.0 \\
NUB-L 
 & 920.2\phantom{*} (+1.5\%) 
 & 299.9 & 3.1 
 & 931.6\phantom{*} (+2.4\%) 
 & 310.6 & 3.0 \\
NUB-Z 
 & 909.6\phantom{*} (+0.3\%) 
 & 300.0 & 3.0 
 & 920.5\phantom{*} (+1.2\%) 
 & 306.8 & 3.0 \\
\midrule
MCA\&NUB-G 
 & 1011.9\phantom{*} (+11.6\%) 
 & 299.9 & 3.4 
 & 1053.7\phantom{*} (+15.8\%) 
 & 351.4 & 3.0 \\
MCA\&NUB-L 
 & 1014.0\phantom{*} (+11.8\%) 
 & 299.9 & 3.4 
 & 1054.5\phantom{*} (+15.9\%) 
 & 351.5 & 3.0 \\
\midrule
\textbf{HOB-Z} 
 & \;\;\textbf{1024.0} \phantom{*}(\textbf{+12.9\%}) $^{\dagger}$
 & 300.0 & \textbf{3.4} 
 & \;\;\textbf{1065.4} \phantom{*}(\textbf{+17.1\%}) $^{\dagger}$
 & 355.1 & 3.0 \\
\bottomrule
\end{tabular}%
}% end resizebox

% \vspace{2pt}
% {\footnotesize 
% $^{*}$ $p < 0.05$,\; $^{**}$ $p < 0.01$,\; $^{***}$ $p < 0.001$ (paired t-test vs.\ MCA\&NUB-L).
% }
\end{table}

\subsection{Online Deployment}
\begin{table}[!t]
  \centering
  \captionsetup{skip=6pt}
  \caption{Online results. ROAS 70\% Rate denotes the fraction of ads whose realized ROAS reaches at least 70\% of the target. $\dagger$ indicates statistical significance under a 1,000-sample nonparametric bootstrap (95\% CI excludes zero).}
  \label{tab:online_res}
  \resizebox{\linewidth}{!}{
  \begin{tabular}{@{}cccccc@{}} % l=left, c=center
    \toprule
     & \textbf{Impression} & \textbf{Click} &\textbf{GMV} & \textbf{Cost} & \textbf{ROAS 70\% Rate}  \\
    \midrule
    \makecell[c]{UB}  & 100\% & 100\% & 100\% & 100\% & 100\% \\
    \makecell[c]{NUB-G} & +2.1\% & +0.9\% & +1.2\% & +0.4\% & -0.3\%\\
    \makecell[c]{NUB-Z} & +1.5\% & +0.6\% & -1.5\% & +1.6\% & +2\%\\
    \makecell[c]{\textbf{HOB-Z}}  & \phantom{*}\textbf{+6.5\%}$^{\dagger}$ & \phantom{*}\textbf{+3.2\%}$^{\dagger}$ & \phantom{*}\textbf{+3.0\%}$^{\dagger}$ &\phantom{*}\textbf{+3.1\%}$^{\dagger}$ &\phantom{*}\textbf{+1\%}$^{\dagger}$ \\
    \bottomrule
  \end{tabular}
  }
\end{table}

\textbf{To answer RQ(4):} We conduct online A/B tests on a large-scale DSP processing billions of daily bid requests, focusing on TargetROAS campaigns. The results show that under the same TargetROAS constraint, our approach enables campaigns to spend more budget while delivering higher GMV. In the FPA-nu channel, we estimate winning prices with a two-tower model: user/ad embeddings are updated hourly. Online serving only computes an inner product to output distribution parameters, and the bid is computed via the Wright $\omega$ function, which is numerically equivalent to the Lambert-W function, adding only 0.53 ms average latency. In the FPA-u channel, the power-law coefficients are refreshed hourly. Then the MCA module is implemented as a simple linear transformation. We use campaign-level randomization with 10 buckets. After a 7-day A/A check, we select the most balanced buckets (max GMV relative difference: 0.46\%) and calibrate the A/B results by the A/A discrepancies. As shown in Table~\ref{tab:online_res}, HOB yields a statistically significant 3.0\% uplift, without compromising the constraints.

\section{Conclusion}\label{conclusion}
In this paper, we study optimal bidding in modern advertising with heterogeneous auction mechanisms and organic-paid coexistence. We propose HOB with two key ideas: (1) marginal cost alignment to balance cost-efficiency across channels, and (2) an effective closed-form bidding strategy for FPA-nu that explicitly accounts for free wins under organic–paid coexistence. Extensive offline experiments and large-scale online A/B tests on a leading DSP validate our approach, delivering a \textbf{3.0\%} GMV improvement in production.

\section{Appendix}
% In your appendix file (e.g., appendix.tex)
\subsection{Proofs}
\label{appendix:proofs}

% =================================================================
% Theorem 1: marginal_cost_equalization
% =================================================================

\subsubsection{Proof 
of Theorem \ref{thm:marginal_cost_equalization}}
\label{mc_eq_proof}
\begin{proof}
Since the denominator in the TargetROAS constraint is strictly positive, the fractional constraint can be rearranged into a linear form. We form the Lagrangian 
$\mathcal{L}$ as:

\begin{equation}
\mathcal{L}({\eta}, \lambda, \gamma) = \sum_{i \in \mathcal{I}} V_i(\eta_i) - \lambda H({\eta}) - \gamma G({\eta})
\end{equation}

where $H(\eta)=\mathrm{R}\cdot{\sum_{i\in\mathcal{I}} C_{i} (\eta_{i})}-\sum_{i\in\mathcal{I}} V_{i} (\eta_{i}) $ denotes the ROAS constraint, and $G(\eta)=\sum_{i\in\mathcal{I}} C_{i} (\eta_{i}) - \mathrm{B}$ denotes the budget constraint. Here, $\lambda$ and $\gamma$ are the Lagrange multipliers associated with the ROAS and budget constraints, respectively.

Under the Linearly Independent Constraint Qualification (LICQ), if $\eta^*=(\eta_{su}^*,\eta_{fu}^*,\eta_{fn}^*)$ is a local optimum, there exist multipliers $(\lambda^*,\gamma^*)$ such that the following KKT conditions hold:

\begin{equation}
H(\eta^*)\le0, \;G(\eta^*)\le0,\; \lambda^*\ge0,\; \gamma^*\ge0,\; \lambda^*H(\eta^*)=0,\; \gamma^*G(\eta^*)=0
\end{equation}

By rearranging the first-order condition $\frac{\partial \mathcal L}{\partial \eta_{i}}
=0$, we have:
\begin{equation}
(1+\lambda^*)\cdot V_i^{'}(\eta_i^{*})
-(\lambda^*\cdot\mathrm{R+\gamma^*})\cdot\,C_i^{'}(\eta_i^{*})
=0
\end{equation}
Assuming $V_i^{'}(\eta_i)>0$ and $C_i^{'}(\eta_i)>0$ (monotonicity of value and cost with respect to the bid control parameter), the ratio of marginal cost to marginal value is given by:
\begin{equation}
\frac{C_i^{'}(\eta_i^{*})}{V_i^{'}(\eta_i^{*})}
= \frac{\lambda^*+1}{\lambda^*\cdot\mathrm{R}+\gamma^*}
\end{equation}
Since the right-hand side depends only on the global multipliers $(\gamma^*, \lambda^*)$ and R, it is constant across all channels 
$i \in \{su,fu,fn\}$. Thus, the MC of each channel is aligned to a unified threshold $\mu$.

\end{proof}

% =================================================================
% Proof of marginal cost in FPA Channel with uniform bidding
% =================================================================

\subsubsection{Proof of MC in uniform FPA Channel}
\label{mc_fu_proof}
\begin{proof}
Fix $\eta_{fu}$ and consider a perturbation $\Delta$ such that $\eta_{fu}+\Delta$ remains in the domain.
Since uniform FPA uses $ b_j=\eta \cdot v_j$ and pays the submitted bid upon winning, the aggregate cost satisfies
\begin{equation}\label{eq:cfpa_def}
C_{fu}(\eta_{fu})=\eta_{fu} \cdot V_{fu}(\eta_{fu}).
\end{equation}
Therefore,

\begin{equation}
\begin{aligned}
    &C_{fu}(\eta_{fu}+\Delta) - C_{fu}(\eta_{fu}) \\
    &\quad = (\eta_{fu}+\Delta) V_{fu}(\eta_{fu}+\Delta) - \eta_{fu} V_{fu}(\eta_{fu}) \\
    &\quad = \eta_{fu} \bigl[ V_{fu}(\eta_{fu}+\Delta) - V_{fu}(\eta_{fu}) \bigr] + \Delta \cdot V_{fu}(\eta_{fu}+\Delta).
    \label{eq:increment_decomp}
\end{aligned}
\end{equation}

% \begin{align}
% C_{fu}(\eta_{fu}+\Delta)-C_{fu}(\eta_{fu})
% &=(\eta_{fu}+\Delta)\cdot V_{fu}(\eta_{fu}+\Delta)-\eta_{fu} \cdot V_{fu}(\eta_{fu}) \nonumber\\
% &=\eta_{fu}\cdot \bigl(V_{fu}(\eta_{fu}+\Delta)-V_{fu}(\eta_{fu})\bigr)
% +\Delta\cdot V_{fu}(\eta_{fu}+\Delta).
% \label{eq:increment_decomp}
% \end{align}
Divide both sides of \eqref{eq:increment_decomp} by $V_{fu}(\eta_{fu}+\Delta)-V_{fu}(\eta_{fu})$ to obtain
\begin{equation}\label{eq:mc_ratio}
\frac{C_{fu}(\eta_{fu}+\Delta)-C_{fu}(\eta_{fu})}{V_{fu}(\eta_{fu}+\Delta)-V_{fu}(\eta_{fu})}
=
\eta_{fu}+\Delta\cdot
\frac{V_{fu}(\eta_{fu}+\Delta)}{V_{fu}(\eta_{fu}+\Delta)-V_{fu}(\eta_{fu})}.
\end{equation}
Rewrite the second term as
\begin{equation}\label{eq:second_term_rewrite}
\Delta\cdot
\frac{V_{fu}(\eta_{fu}+\Delta)}{V_{fu}(\eta_{fu}+\Delta)-V_{fu}(\eta_{fu})}
=
\frac{V_{fu}(\eta_{fu}+\Delta)}{\frac{V_{fu}(\eta_{fu}+\Delta)-V_{fu}(\eta_{fu})}{\Delta}}.
\end{equation}
Taking $\Delta\to 0$ and using differentiability of $V$ at $\eta_{fu}$ yields
\begin{equation}
\lim_{\Delta\to 0}
\frac{V_{fu}(\eta_{fu}+\Delta)}{\frac{V_{fu}(\eta_{fu}+\Delta)-V_{fu}(\eta_{fu})}{\Delta}}
=
\frac{V_{fu}(\eta_{fu})}{V_{fu}'(\eta_{fu})}.
\end{equation}
Applying the limit to \eqref{eq:mc_ratio} gives
\begin{equation}
MC_{fu}(\eta_{fu})
=
\eta_{fu}+\frac{V_{fu}(\eta_{fu})}{V_{fu}'(\eta_{fu})}.
\end{equation}
\end{proof}

% =================================================================
% Proof of Theorem 2:marginal cost in FPA Channel with non-uniform bidding
% =================================================================

\subsubsection{Proof of Theorem\ref{thm:marginal_cost}}
\label{mc_fn_proof}
\begin{proof}
The proof hinges on the first-order condition (FOC) from the surplus maximization problem and the application of the chain rule to the aggregate cost and value functions.
With a given control parameter $\eta_{fn}$, the optimal bid $b_{fn,j}(\eta_{fn})$ for impression $j$ solves~\eqref{eq:surplus_continuous}. 
Define the expected surplus objective as
\[
S(b) \triangleq \bigl(\eta_{fn} \cdot v_{fn,j}-b\bigr)\,\cdot p_{fn,j}(b).
\]
For impressions with an interior optimum $b_{fn,j}(\eta_{fn})>0$, the necessary first-order condition is obtained by differentiating $S(b)$ with respect to $b$ and setting the derivative to zero:
\[
\frac{\partial S(b)}{\partial b}
= -p_{fn,j}(b) + \bigl(\eta_{fn}\cdot v_{fn,j}-b\bigr)\,\cdot p'_{fn,j}(b)=0,
\]
which can be rearranged as
\begin{equation}
\label{eq:foc_sketch}
p_{fn,j}\!\bigl(b_{fn,j}\bigr) + b_{fn,j}\,\cdot p'_{fn,j}\!\bigl(b_{fn,j}\bigr)
= \eta_{fn}\,\cdot v_{fn,j}\,\cdot p'_{fn,j}\!\bigl(b_{fn,j}\bigr).
\end{equation}

This condition links the optimal bid to the value and the control parameter $\eta_{fn}$. The MC is defined as $MC_{fn}(\eta_{fn}) = \frac{C^{\prime}_{fn}(\eta_{fn})}{V^{\prime}_{fn}(\eta_{fn})}$. Under the ex-ante formulation in Section \ref{Optimal Bid FPA nub}, the aggregate value and cost are differentiable, avoiding the non-differentiability of replay-style indicators.

\begin{enumerate}
    \item Derivative of Total Value ($V_{fn}'(\eta_{fn})$): Applying the chain rule to $V_{fn}(\eta_{fn}) = \sum_{j \in \mathcal{J}_{fn}} v_{fn,j} \cdot p_{fn,j}(b_{fn,j}(\eta_{fn}))$ gives:
    \begin{equation}
    \label{eq:value_deriv_sketch}
    V_{fn}'(\eta_{fn}) = \sum_{j \in \mathcal{J}_{fn}} v_{fn,j} \cdot p'_{fn,j}(b_{fn,j}) \cdot \frac{d(b_{fn,j})}{d\eta_{fn}}.
    \end{equation}

    \item Derivative of Total Cost ($C_{fn}'(\eta_{fn})$): Applying the chain rule to $C_{fn}(\eta_{fn}) = \sum_{j \in \mathcal{J}_{fn}} b_{fn,j}(\eta_{fn}) \cdot p_{fn,j}(b_{fn,j}(\eta_{fn}))$ and replacing the result using the FOC from Equation~\eqref{eq:foc_sketch} yields:
    \begin{equation}
    \label{eq:cost_deriv_sketch}
    C_{fn}'(\eta_{fn}) = \sum_{j \in \mathcal{J}_{fn}} v_{fn,j} \cdot \eta_{fn} \cdot p'_{fn,j}(b_{fn,j}) \cdot \frac{d(b_{fn,j})}{d\eta_{fn}}.
    \end{equation}
\end{enumerate}

Inspecting, we observe that $C_{fn}'(\eta_{fn}) = \eta_{fn} \cdot V_{fn}'(\eta_{fn})$. Taking the ratio of the two derivatives, the summation terms cancel out:
\begin{equation}
\begin{aligned}
MC_{fn}(\eta_{fn}) &= \frac{C_{fn}'(\eta_{fn})}{V_{fn}'(\eta_{fn})} \\
 &= \frac{\eta_{fn} \cdot \sum_{j \in \mathcal{J}_{fn}} \left( v_{fn,j} \cdot p'_{fn,j}(b_{fn,j}) \cdot \frac{d(b_{fn,j})}{d\eta_{fn}} \right)}{\sum_{j \in \mathcal{J}_{fn}} \left( v_{fn,j} \cdot p'_{fn,j}(b_{fn,j}) \cdot \frac{d(b_{fn,j})}{d\eta_{fn}} \right)} = \eta_{fn}.
\end{aligned}
\end{equation}

This establishes that the MC of value acquisition with respect to the control parameter $\eta_{fn}$ coincides with $\eta_{fn}$ itself.
\end{proof}

\subsubsection{Proof of Theorem~\ref{thm:exist_uni_mu}}
\label{opt_u_proof}
\begin{proof}
Consider a multi-channel setting where the aligned bidding control parameter $\mu$ jointly determines the total value $V(\mu)$ and total cost $C(\mu)$. The following properties are directly implied by the per-channel definitions.

\begin{proposition}[Monotonicity]
The following properties hold for the aggregated market environment:
\begin{itemize}
    \item \textbf{Monotonicity of value.} By construction of the channel-wise value functions, the total value $V(\mu)$ is monotonically non-decreasing in $\mu$, with range $[0,+\infty)$.
    \item \textbf{Monotonicity of cost.} By construction of the channel-wise cost functions, the total cost $C(\mu)$ is monotonically non-decreasing in $\mu$, with range $[0,+\infty)$.
    \item \textbf{Monotonicity of ROAS.} Under MC alignment, the MC satisfies
    $MC(\mu) = \mu$, hence it is monotonically increasing in $\mu$ for each channel and for overall. Consequently, the ROAS function is monotonically non-increasing in $\mu$, with range $[0, ROAS_{\max}]$, where $ROAS_{\max}$ denotes the maximum attainable ROAS of a single best impression in the impressions pool.
\end{itemize}
\end{proposition}

\paragraph{Existence and uniqueness.}
By the intermediate value theorem and the monotonicity above, the feasible solution exists and is unique when the feasible frontier is strictly monotone.

\begin{itemize}
\item \textbf{Only consider budget constraint:}
Assume a finite set of candidate impressions indexed by $k=1,\dots,K$, each with value $v_k$, cost $c_k$, and threshold $\mu_k$ (i.e., impression $k$ is selected when $\mu\ge \mu_k$). For any $\mathrm{B}>0$, define
\begin{equation}
\mu_B^*
= \max \left\{ \mu \ \middle|\ 
\sum_{k=1}^{K} c_k \cdot \mathbb{I}(\mu \geq \mu_k) \leq \mathrm{B}
\right\}.
\label{eq:mu_budget}
\end{equation}

\item \textbf{Only consider ROAS constraint:}
For any target ROAS level $\mathrm{R}\in[0,ROAS_{\max}]$, define
\begin{equation}
\mu_R^*
= \max \left\{ \mu \ \middle|\ 
\frac{\sum_{k=1}^{K} v_k \cdot \mathbb{I}(\mu \geq \mu_k)}
{\sum_{k=1}^{K} c_k \cdot \mathbb{I}(\mu \geq \mu_k)}
\geq \mathrm{R}
\right\}.
\label{eq:mu_roas}
\end{equation}
\end{itemize}

\paragraph{Optimal solution.}
The overall feasible solution is given by
\begin{equation}
\mu^*=\min\!\left(\mu_B^*,\mu_R^*\right).
\label{eq:mu_star}
\end{equation}
Moreover, since $V(\mu)$ is monotonically non-decreasing, $\mu^*$ is optimal and coincides with the largest feasible threshold under the constraints. In practice, given the monotone relationship between $\mu$ and the constraints, $\mu^*$ can be obtained either by constraint inversion (solving \eqref{eq:mu_budget}--\eqref{eq:mu_roas}) or by online feedback control that dynamically adjusts $\mu$ to drive the system toward the optimal point.
\end{proof}

\subsection{Pseudo-code}
\label{appendix:Pseudo-code}
% The Winning-Price Model training is presented in Algorithm~\ref{alg:training_simplified}.

% \begin{algorithm}[htbp]
%     \caption{Offline ZIE Distribution Model training}
%     \label{alg:training_simplified}

%     \SetAlgoLined
%     \SetKwInOut{Input}{Input}
%     \SetKwInOut{Output}{Output}
%     \SetKwFunction{ZIENLL}{ZIENLL}
%     \SetKwProg{Fn}{Function}{:}{}
    
%     \Input{
%         Training dataset $\mathcal{D} = \{(\mathbf{x}_i, wp_i)\}_{i=1}^N$
%     }
%     \Output{
%         A trained distribution prediction model, $\text{DistModel}$
%     }
%     \BlankLine
    
%     % --- Main Logic ---
%     Initialize model parameters $\theta$ and an optimizer (e.g., Adam) \\
%     \For{each epoch}{
%         \For{each batch $\{(\mathbf{x}_j, wp_j)\}_{j\in \mathcal{J}} \subset \mathcal{D}$}{
            
%             $(\hat{\pi}_j, \hat{\lambda}_j) \leftarrow \text{DistModel}(\mathbf{x}_j; \theta)$ \\
            
%             $\mathcal{L} \leftarrow \frac{1}{\lvert \mathcal{J}\rvert} \sum_{j\in \mathcal{J}} \ZIENLL(\hat{\pi}_j, \hat{\lambda}_j, wp_j)$ \\

%             Perform a gradient descent step on $\mathcal{L}$ to update $\theta$
%         }
%     }
%     \KwRet{$\text{DistModel}$}
    
%     \BlankLine
%     \hrule
%     \BlankLine

%     \Fn{\ZIENLL($\hat{\pi}, \hat{\lambda}, wp$)}{
%         \tcp{Calculates the Negative Log-Likelihood for a single sample}
%         \eIf{$wp = 0$}{
%              \KwRet $-\log(\hat{\pi})$
%         }{
%              \KwRet $-\log(1 - \hat{\pi}) - \log(\hat{\lambda}) + \hat{\lambda} wp$
%         }
%     }
% \end{algorithm}

\begin{algorithm}[htbp]
\DontPrintSemicolon
\SetKwInOut{KwIn}{Input}
\SetKwInOut{KwOut}{Output}
\SetKwFunction{WOBid}{WOBid}
\SetKwProg{Fn}{Function}{:}{}

\caption{HOB: MC-Aligned Bidding under Heterogeneous Channels}
\label{alg:hob}

\KwIn{log $\mathcal{D}=\{(x_j,\mathrm{wp}_j)\}$;\ request stream $\mathcal{R}$ split into
      $\{\mathrm{SPA}_u,\mathrm{FPA}_u,\mathrm{FPA}_{nu}\}$;\ 
      power-law fit $V_{fu}(\eta)=a(\eta+c)^b$;\ target $=$ budget $B$ \textbf{or} ROAS $R$;\ horizon $T$;\ PID gains $(K_p,K_i,K_d)$}
\KwOut{online bid stream $\{b_j\}_{j\in\mathcal{R}}$ with slot-wise updated $\mu_t$ and $(\eta_{\mathrm{su},t},\eta_{\mathrm{fu},t},\eta_{\mathrm{fn},t})$}

\tcp{\emph{(i) Offline: ZIE parameter learning for FPA-NU}}
train $f_\theta(x)\!\to\!(\pi,\lambda)=(\sigma(h_\pi),\mathrm{softplus}(h_\lambda))$ by $\min_\theta \sum_j \ell_j$% with\;
\quad$\ell_j=\mathbf{1}[\mathrm{wp}_j{=}0](-\log\pi_j)+\mathbf{1}[\mathrm{wp}_j{>}0](-\log(1{-}\pi_j)-\log\lambda_j+\lambda_j\mathrm{wp}_j)$\;

\tcp{\emph{(ii) Online: PID-control $\mu$ to track ROAS / budget-pacing target}}
$\mu\gets\mu_0$;\ \ $I\gets 0$;\ \ $e_{\mathrm{prev}}\gets 0$;\ \ $(V_0,C_0)\gets(0,0)$\;
\ForEach{control step $t=1,\ldots,T$ with pacing fraction $\tau_t=t/T$}{
  $\eta_{\mathrm{su}},\eta_{\mathrm{fn}}\gets\mu$;\ \ $\eta_{\mathrm{fu}}\gets(b\mu-c)/(b+1)$\;
  \ForEach{$j$ in slot-$t$ requests with value $v_j$}{
    $b_j\gets
      \begin{cases}
        \eta_{\mathrm{su}} v_j & j\in\mathrm{SPA}_u \\
        \eta_{\mathrm{fu}} v_j & j\in\mathrm{FPA}_u \\
        \WOBid(\eta_{\mathrm{fn}},v_j,\pi_j,\lambda_j) & j\in\mathrm{FPA}_{nu}
      \end{cases}$\;
    \lIf{$b_j\ge\mathrm{wp}_j$}{$V_t\mathrel{+}=v_j$;\ $C_t\mathrel{+}=(\,\mathrm{wp}_j\text{ if SPA else }b_j\,)$}
  }
  \lIf{budget mode}{$e_t\gets C_t/(B\tau_t)-1$}\lElse{$e_t\gets R-V_t/\max(C_t,\varepsilon)$}%\tcp*[r]{$e_t>0\Rightarrow\mu\!\uparrow$}
  $I\gets I+e_t$;\ \ $D\gets e_t-e_{\mathrm{prev}}$;\ \ $e_{\mathrm{prev}}\gets e_t$\;
  $\mu\gets\mathrm{clip}(\mu+K_p e_t+K_i I+K_d D,\ 0,\ \mu_{\max})$\;
}

\Fn{\WOBid{$\eta,v,\pi,\lambda$}}{
  \lIf{$(1{-}\pi)(1{+}\lambda\eta v)\le 1$}{\Return $0$}%\tcp*[r]{surplus non-increasing on $[0,\eta v]$}
  $\alpha\gets\exp(1+\lambda\eta v)\,/\,(1-\pi)$\tcp*[r]{Lambert-$W$ argument}
  $w\gets W_0(\alpha)$\tcp*[r]{principal branch, e.g.\ \texttt{scipy.special.lambertw}}
  $b^\star\gets\eta v+(1-w)/\lambda$\;
  \Return $\mathrm{clip}(b^\star,\ 0,\ \eta v)$\;
}
\end{algorithm}

\section{GenAI Usage Disclosure}
The authors employed LLMs solely for post-writing refinement and language polishing. All substantive components of this work, including research conception, methodological design, theoretical analysis, experimental execution, and interpretation of results, were independently conceived and carried out by the authors.
% During the preparation, the authors used ChatGPT for language editing and proofreading to improve the readability of the manuscript. The authors reviewed and edited all AI-generated suggestions and take full responsibility for the content of this publication.
\bibliographystyle{ACM-Reference-Format}
\bibliography{ref}

% \section*{Presenter Bio}
% \textbf{Qi Li} is a Senior Algorithm Engineer at Alibaba Group, working on large-scale auction and auto-bidding systems for e-commerce advertising. His work focuses on Real-Time Bidding (RTB) and computational advertising, including bidding strategy design, budget pacing, and ROAS optimization under heterogeneous auction mechanisms. He has contributed to the design and deployment of the HOB (Holistically Optimized Bidding) framework in Alibaba’s display advertising platforms, with a focus on improving bidding efficiency and advertiser performance. His research interests include real-time bidding, auction mechanism design, and integrating foundation models into industrial ad delivery and optimization workflows. He will be the main presenter for this work.

\end{document}